\documentclass[a4paper]{iopart}

\usepackage{iopams}  
\usepackage{bm}
\usepackage{graphicx}

\begin{document}

\title[Covariance analysis of finite temperature density functional theory]{Covariance analysis of finite temperature density functional theory: symmetric nuclear matter}

\author{A Rios$^1$ \& X Roca-Maza$^2$}

\address{$^1$Department of Physics, Faculty of Engineering and Physical Sciences, University of Surrey, Guildford, Surrey GU2 7XH, United Kingdom}
\address{$^2$Dipartimento di Fisica, Universit{\`a} degli Studi di Milano and INFN, Sezione di
Milano, 20133 Milano, Italy}
\ead{a.rios@surrey.ac.uk}
\vspace{10pt}

\begin{abstract}
We study symmetric nuclear matter at finite temperature, with particular emphasis on the liquid-gas phase transition. We use a standard covariance analysis to propagate statistical uncertainties from the density functional to the thermodynamic properties. We use four functionals with known covariance matrices to obtain as wide a set of results as possible. Our findings suggest that thermodynamical properties are very well constrained by fitting data at zero temperature. The propagated statistical errors in the liquid-gas phase transition parameters are relatively small.
\end{abstract}

\section{Introduction}

Traditional nuclear physics calculations have often ignored error estimates. A series of recent efforts in both energy density functional \cite{Reinhard2010,Dobaczewski2014} and ab initio calculations \cite{Hebeler2010,Hagen2013} have tried to remedy this deficiency. In this special issue, several contributions discuss the study, propagation and understanding of statistical and systematic errors in nuclear theory calculations. Here, we will not discuss so much error propagation techniques, but rather concentrate on the impact of statistical errors in finite temperature calculations. This is an exploratory study, aimed at answering a few key questions. Among others, we want to discuss the significant issue of whether finite temperature results, which are generally ignored in fitting protocols, are reliably constrained by their zero temperature counterparts. 

By its very nature, most nuclear theory calculations involve some sort of quantum many-body calculation. There are a variety of error sources in such many-body predictions, and these errors are frequently intertwined. Generally, one can divide error sources into numerical uncertainties, approximations in the many-body method, systematic errors associated to the choice of the nuclear effective interaction or energy density functional (EDF) and statistical errors -- evaluated from the so-called covariance analysis -- that originate from the employed fitting protocol of a given model. Some of these uncertainties, i.e. particularly those of a numerical and a many-body nature, are the same as in other fields of physics \cite{PRA2011}. 

Systematic errors in the many-body approximations can be hard to tackle. Within EDF, characterising the error of the density functional is not necessarily an easy task \cite{Klupfel2009,RocaMaza2011,RocaMaza2012}. Here, we do not discuss this issue explicitly. We work, however, with four different density functionals to have a wide subset of theoretical results that can guide our understanding. 

Uncertainties of the numerical methods on many-body simulations are also a source of theoretical error. The accuracy of numerical simulations, however, can often be tested by a variety of means, particularly if  parallel calculations performed with other methods are available. Finite temperature calculations for infinite matter can nowadays be performed extremely quickly and to a large degree of accuracy, to the extent that numerics are not an issue \cite{Rios10}. 

In addition to any potential numerical or many-body bias, a major (if not the largest) source of uncertainty in nuclear physics is the underlying effective interaction (or EDF). The non-perturbative nature of QCD at low energies essentially precludes the construction of a nucleon-nucleon interaction. One therefore relies on parametrized interactions, which in some cases can be derived from systematic expansions \cite{Bogner2010}. Some of these underlying interactions, like phase-shift-equivalent interactions, can be extremely sophisticated, depending on a wide range of parameters and physical processes. Error propagation from these realistic interactions into physical observables has only been studied recently \cite{Hagen2013,Navarro2014,Navarro2014b}. Other parametrizations, like Skyrme energy density functionals, provide relatively simple phenomenological descriptions of the nuclear interaction in the medium. Because of their accessible nature, EDF calculations allow for large-scale calculations over the whole nuclear chart \cite{Erler2012}. A key issue, though, is if and how to reduce the uncertainties associated to the EDF parametrizations themselves. 

In the following, we will not discuss either numerical or many-body errors. We will restrict ourselves to the propagation of statistical errors from the EDF into finite temperature calculations of nuclear matter. We note also that we will not discuss specifically any propagation error method. Instead, we base our analysis on a standard propagation technique, assuming that the objective function is a $\chi^2$-distribution of parameters which follows a parabolic behaviour around the minimum \cite{Reinhard2010,RocaMaza2014}. 
While these assumptions might not be entirely correct in nuclear EDFs, the relatively simple covariance analysis that arises already provides a powerful analysis technique, relevant for the present exploratory study. 

Temperature offers an additional degree of freedom with critical information. In particular, the covariance analysis can help identify which, if any, of the zero temperature fit properties determines the temperature dependence of nuclear observables.  EDFs are generally fitted to zero temperature properties, and the extent to which these are able to constrain the temperature dependence of the functional is relatively unexplored. As discussed by one of us in Ref.~\cite{Rios10}, several correlations exist between zero temperature and finite temperature properties of nuclear matter. If these correlations are due to the underlying functional, one may also expect to see them as statistical correlations after error propagation from the functional to the thermodynamical properties. Such correlations are relevant also for finite temperature calculations of finite nuclei \cite{Schunck2014}.

The existence of the liquid-gas phase transition, which has been studied extensively with a variety of experiments \cite{pochodzalla95,pochodzalla97,natowitz02,elliott02,natowitz02b,borderie08}, can provide additional constraints on the nuclear matter equation of state (EoS). Extrapolations of this data to infinite nuclear matter have recently been attempted \cite{Elliott2013}. Independently of the biases underlying such studies, one would like to know the precision with which liquid-gas phase transition parameters can be determined theoretically and how these errors compare to experimental estimates. The critical properties of the liquid-gas transition also provide a characterisation of the thermal properties of nuclei. We will focus our study on the thermodynamical properties of infinite symmetric matter. Note, however, that error propagation in hot isospin asymmetric nuclear matter can provide an additional handle on the symmetry energy \cite{tsang01,sfienti09}. 

\section{Results}

\subsection{Equation of state}

We perform calculations of finite temperature Skyrme density functionals for infinite matter. Details of the extension of EDF calculations to finite temperature can be found elsewhere \cite{Rios10,vautherin96}. Within Skyrme EDF, the chemical potential, $\mu$, is fixed at every density, $\rho$, and temperature, $T$, via the normalisation of the momentum distribution, 
\begin{equation}
\rho = 4 \int \frac{\textrm{d}^3 k}{(2 \pi)^3} f(k) \, , 
\end{equation}
which is a Fermi-Dirac distribution, 
\begin{equation}
f(k) = \left[ 1+ e^{\left( \frac{\hbar ^2 k^2}{2m^*} + U - \mu \right)/T} \right]^{-1} \, .
\label{eq:FD} 
\end{equation}
The single-particle potential is formed of two basic ingredients. On the one hand, the effective mass,
\begin{equation}
\frac{m^*(\rho)}{m}= \frac{1}{ 1 + \frac{2m}{\hbar^2} C_0^\tau \rho  } \, ,
\label{eq:effmass}
\end{equation}
encodes all the momentum dependence of the mean-field. On the other, the single-particle shift,
\begin{equation}
\bar U  = 2 C_{00}^{\rho \rho} \rho + (2+\alpha) C_{0D}^{\rho \rho} \rho^{\alpha+1} \, ,
\end{equation}
includes the rearrangement term. We use the notation for EDF parametrizations of Ref.~\cite{Kortelainen2010}. Having found the chemical potential $\mu$, the distribution $f(k)$ is fixed and one can compute the bulk thermodynamic properties: energy, entropy, free energy and pressure \cite{vautherin96}.  

Using covariance analysis, we want to study the uncertainty of thermodynamical properties associated to the underlying parameters of the Skyrme EDF. To this end, we will use 4 parametrizations of non-relativistic functionals: SLy5min \cite{RocaMaza2014}, UNEDF0 \cite{Kortelainen2010}, UNEDF1 \cite{Kortelainen2012} and UNEDF2 \cite{Kortelainen2014}. We choose these functionals because their fit covariance matrices are published. SLy5min has been obtained by one of us \cite{RocaMaza2014} in an effort to reproduce the original Skyrme Lyon 5 parametrization \cite{Chabanat1998}. This was fitted using the original set of data, which includes (a) saturation properties of symmetric matter, (b) a realistic EoS for neutron matter, (c) the binding energies (radii) of 5 (4) nuclei and (d) a spin-orbit term fitted to the $3p$ neutron splitting in $^{208}$Pb. The following relations are needed to translate the original Skyrme force parameters into EDF constants:
\begin{eqnarray}
C_{00}^{\rho \rho} = \frac{3}{8} t_0 \, , \\ 
C_{0D}^{\rho \rho} = \frac{1}{16} t_3\, ,  \\
C_0^\tau = \frac{1}{16} \left[ 3t_1 + 5t_2 + 4 t_2 x_2 \right] \, .
\end{eqnarray}
We note that these three isoscalar parameters, together with the density exponent $\alpha$, completely characterise the zero and finite temperature symmetric matter EoS of Skyrme functionals. 

The UNEDF family of functionals are state-of-the-art Skyrme parametrizations in the context of Hartree-Fock-Bogolioubov theory. They are fitted by means of derivative-free optimisation algorithms, that provide reliable object function minimizations \cite{Kortelainen2010}. The data they are fitted to includes a variety of finite nuclei and isoscalar and isovector bulk properties. In terms of bulk nuclear matter, rather than fitting the EDF parameters, one imposes a series of \emph{physical} properties (saturation density and energy; compressibility; symmetry energy in the isovector channel) and then reconstructs the corresponding $C_i^x$ EDF constants. Some of these parameters, like the saturation density, are left to vary within a given range and hence carry their own statistical uncertainty. Other parameters, like the exponent $\alpha$, are fixed from the start. 

The only parameters required in this study are those of an isoscalar, bulk nature. In consequence, there is a small number (typically 2 or 3) of parameters that carry relevant statistical information for error propagation purposes. The original parametrization UNEDF0 \cite{Kortelainen2010} allowed for changes in both the saturation density, $\rho_0$, and energy, $e_0$, whereas the incompressibility, $K_0$, and the effective mass, $m_0$, were fixed to their extremal values. In UNEDF1 \cite{Kortelainen2012}, instead, the saturation energy and the incompressibility saturated their bounds and hence were fixed. In contrast, errors in the effective mass, and hence $C^\tau_0$, as well as the saturation density, were introduced. Finally, UNEDF2 \cite{Kortelainen2014} was fitted taking shell effects into consideration. For this parametrization, in addition to $\rho_0$ and $m^{*}_0$, the incompressibility was constrained by the data. These differences are useful for the following analysis, as they provide an indication of which parameters are relevant for finite temperature properties.

The central values and standard deviations for the EDF used in this work were taken from the respective original publications. To carry out the covariance analysis, we have performed calculations at the central values as well as at the two extremes, $p_i \pm \sigma_{p_i}/10$, of each free parameter, $p_i$, in the functional. $\sigma_{p_i}$ is the standard deviation of parameter $p_i$. As discussed in Ref.~\cite{RocaMaza2014}, this provides a way to compute the step size for the numerical derivatives of the objective function, $\partial_{p_i} \partial_{p_j}\chi^2({\bf p})$, that correspond to the curvature matrix. We have also checked that a change of these extreme values provides very similar results, which confirms the quality of the step size. The original statistical correlations have been propagated to the different quantities of interest. All the errors presented in the following have been computed at $1\sigma$.

\begin{figure}[t]
\begin{center}
\includegraphics[width=0.7\linewidth]{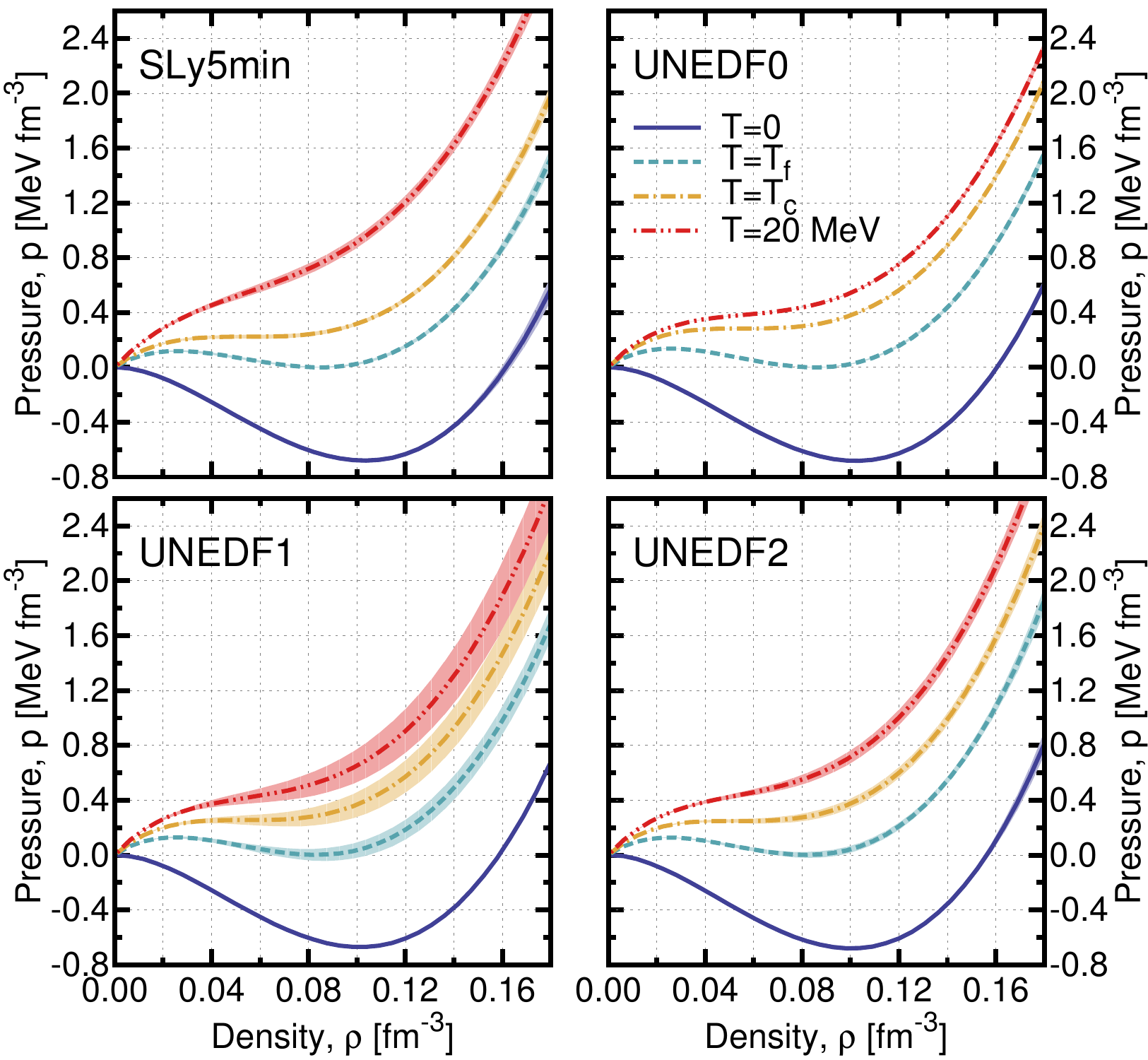}
\end{center}
\caption{(Color online) Pressure isotherms for various temperatures: $T=0$ (solid line), $T=T_f$ (dashed line), $T=T_c$ (dashed-dotted line) and $T=20$ MeV (dash-double-dotted line). The bands correspond to $1 \sigma$ statistical uncertainties propagated as described in the text. In some cases, like UNEDF0, the errors are smaller than the width of the lines. }
\label{fig:EoS}
\end{figure}

With a given Skyrme parametrization, one can compute the EoS of symmetric matter in the form $p(\rho,T)$. Having access to the standard deviations and covariance matrices of the underlying parameters, we can also find the statistical uncertainties of the pressure for a fixed density and several temperatures. We present the results of this exercise in \Fref {fig:EoS}. For every EDF, we show the EoS for 4 fixed temperatures. The solid line represents the zero-temperature result. As expected, the pressure is zero both at the origin, $\rho=0$, but also at the saturation point, $\rho_0 \approx 0.16$ fm$^{-3}$. The width of the colour bands in  \Fref {fig:EoS} is the standard deviation of each pressure and, at saturation, this is very small. The saturation density is well constrained, to better than $0.002$ fm$^{-3}$ by all these functionals. Incidentally, the fact that $p(\rho_0,0)=0$ is imposed also as a  fit property for all the functionals. It is therefore not surprising that the errors around the saturation point are extremely small. 

As temperature increases, the region of negative pressures decreases until, at one point, the pressure becomes a positive definite function. The temperature at which this occurs is the so-called flashing temperature, $T_f$. We show with a dashed line in \Fref {fig:EoS} the EoS at the flashing temperature for the four functionals. The flashing point is also characterised by a density, $\rho_f$, the single point (other than the origin) that satisfies the constraints:
\begin{equation}
p = \frac{\partial p}{\partial \rho} = 0 \, .
\end{equation}
The flashing point is relevant, because it provides the maximum temperature at which the system can be self-bound - that is, at which it can be at zero pressure. Above this temperature, isothermal nuclear matter would in principle expand due to its own positive pressure. In other words, the flashing temperature is an upper limit for the maximal, limiting temperature that a nucleus can withstand \cite{natowitz02,baldo99}.

The flashing isotherm is very similar for all parametrizations. We present in Table~\ref{table} the results for the flashing temperature and the flashing density for the four EDFs (we will comment upon these later). The temperature $T_f$ is within $11-13$ MeV and the density  $\rho_f$ is slightly above half saturation, as in most modern Skyrme parametrizations \cite{Rios10}. We observe that, in UNEDF1 and UNEDF2, the error band for the flashing isotherm has increased with respect to the zero-temperature one, particularly at densities above saturation. This is natural, in the sense that only zero-temperature, sub-saturation properties have been fixed in the fitting procedure. One therefore expects that as temperature increases, and the results move further away from $T=0$, the error bars in the EoS increase. 

These expectations are confirmed by the critical isotherm, displayed in \Fref {fig:EoS} with a dash-dotted line. At this temperature, the pressure is positive throughout the density regime. The critical point is found by requiring that only one phase is available,
\begin{equation}
\frac{\partial p}{\partial \rho}  =  \frac{\partial^2 p}{\partial \rho^2} = 0 \, .
\end{equation}
In other words, the pressure becomes a monotonous and concave function of density above the critical isotherm, $T=T_c$. 
Again, the critical properties of these EDFs are reported in Table~\ref{table} and will be discussed further on. Let us note here that the properties of the critical isotherm are rather standard \cite{Rios10}. Finally, we show for completeness an isotherm above the critical point, $T=20$ MeV. This corresponds to the double-dotted-dashed line in \Fref {fig:EoS}. 

We have further investigated the density and temperature dependence of the statistical uncertainties in the EoS. We do not provide explicit figures for these properties for brevity, but we can confirm the expected trends. The zero temperature pressure is best constrained, i.e. it shows the smallest errors, below and around saturation.  As density increases above saturation, the errors in the EoS increase for all parametrizations. 

The temperature dependence of the errors is parametrization-dependent, though. UNEDF1 and UNEDF2 both have a noticeable increase of the errors as a function of temperature. The temperature dependence of the SLy5-min EoS errors is, in contrast, rather mild. UNEDF0 is rather unique, in that it displays completely temperature-independent errors. We attribute this somewhat anti-intuitive result to the fact that the isoscalar effective mass is fixed in UNEDF0. $m^{*}_0$ determines to a large extent the temperature dependence of several thermodynamical properties \cite{Rios10}.  Since this parameter is fixed in the fitting procedure, it has no statistical error and the uncertainties hence become insensitive to temperature - even though they depend on density. In contrast, the functional with the more uncertain $m^{*}_0$, UNEDF1, shows the largest temperature dependence in the statistical uncertainties. Future EDF parametrizations could take this into account. If finite temperature properties need to be constrained for any specific reason, the effective mass should be determined to a large accuracy. 

\subsection{Liquid-gas phase transition}

\begin{table}[t!]
\caption{Critical (above) and flashing (below) temperatures and densities for the 4 EDF parametrizations of interest. In column 4, we also provide the critical pressure (above) and the maximum latent heat (below). For comparison, we provide the experimentally derived results of the critical point of Ref.~\cite{Elliott2013}.}
\label{table}
\begin{indented}
\item[]\begin{tabular}{lccc}
\br
Functional  & $T_c$  [MeV]  & $\rho_c$ [fm$^{-3}$] & $p_c$   [MeV fm$^{-3}$] \\
\mr
SLy5min  & $14.6 \pm 0.1$ &  $0.055 \pm 0.001$ & $0.22 \pm 0.01$ \\
UNEDF0  & $17.99 \pm 0.03$ &  $0.0557 \pm 0.0003$ & $0.283 \pm 0.002$ \\
UNEDF1  & $16.8 \pm 1.2$ &  $0.053 \pm 0.002$ & $0.26 \pm 0.03$ \\
UNEDF2  & $16.4 \pm 0.5$ &  $0.0526 \pm 0.0005$ & $0.24 \pm 0.01$ \\
Ref. \cite{Elliott2013}  & $17.9 \pm 0.4$ &  $0.06 \pm 0.01$ & $0.31 \pm 0.07$ \\
\mr
$\phantom{a}$ & $T_f$ [MeV] & $ \rho_f $ [fm$^{-3}$] & $L_H$ [MeV] \\
\mr
SLy5min  & $11.74 \pm 0.06$ & $0.085 \pm 0.001$ & $29.8 \pm 0.4$  \\
UNEDF0  & $13.50 \pm 0.02$ & $0.0854 \pm 0.0005$ & $29.59 \pm 0.07$  \\
UNEDF1  & $12.9 \pm 0.6$ & $0.083 \pm 0.001$ & $29.26 \pm 0.03$  \\
UNEDF2  & $12.7 \pm 0.3$ & $0.0818 \pm 0.0004$ & $29.5 \pm 0.2$  \\
\br
\end{tabular}
\end{indented}
\end{table}

Having verified that our error propagation strategy works for the EoS, we discuss explicitly the liquid-gas phase transition. We note that, to this end, we have performed calculations of the critical and flashing point for every parameter set. We show a summary of the liquid-gas properties and their associated statistical uncertainties in Table~\ref{table}. The first 4 rows of the table provide the information associated to the critical point for the functionals of interest. We find that, except for UNEDF1, all critical temperatures are constrained to better than $0.5$ MeV. In relative terms, this corresponds to uncertainties to within a few percent of the central value. Critical densities are even more constrained, to better than $4 \, \%$ in relative terms. The critical pressure also shows extremely small statistical uncertainties. In all cases, the differences between results of different EDFs are larger than the statistical errors. 

The dimensionless parameter, $\gamma_c = \frac{p_c}{\rho_c T_c}$ is close to the value $\gamma_c \approx 0.28$ in all cases, which is typical of most Skyrme parametrizations. In general, these results agree with those associated to most modern Skyrme forces \cite{Rios10}. We provide, for completeness, the critical point values obtained from a recent empirical analysis based on a series of experimental results \cite{Elliott2013}. For both the critical density and pressure, the experimental values fall within the theoretical predictions and have larger errors bars. The empirical critical temperature, in contrast, only agrees with (a) the UNEDF0 value which, as discussed earlier, has the smallest theoretical error due to the strong constraint on the isoscalar effective mass and (b) the UNEDF1 prediction, due to its large error bar. 

The lower section of Table~\ref{table} concentrates on the flashing properties of nuclear matter. The flashing temperature is constrained to a similar accuracy as the critical temperature. The range $T_f = 11.7-13.5$ MeV, allowed by our theoretical calculations, is somewhat smaller than the large range obtained with a wide variety of Skyrme functionals observed in Ref.~\cite{Rios10}, which suggests a systematic error dominance in this observable. We note, however, that modern forces tend to fall within an analogously narrow flashing temperature region. Similar conclusions hold for the flashing density, which is predicted to lie within a narrow window.

Finally, we provide in the last column of the lower section of Table~\ref{table} the maximum value of the latent heat of the liquid-gas phase transition. As discussed in Ref.~\cite{carbone11}, the latent heat provides a further characterisation of the  phase transition which, moreover, can be accessed experimentally \cite{borderie08,bonnet09}. The latent heat can be computed at each temperature by taking the difference between the gas and the liquid entropies along the coexistence line:
\begin{equation}
L(T) = T \left[ s_g(T) - s_l(T) \right] \, .
\label{eq:latent}
\end{equation}
At low temperatures, the latent heat tends to the saturation energy and has a universal slope \cite{carbone11}:
\begin{equation}
L(T) \approx e_0 + \frac{5}{2} T  \, .
\end{equation}
Because the coexistence zone shrinks as temperature increases, the latent heat is zero at the critical point, $L(T_c)=0$. In consequence, the latent heat as a function of temperature must present a maximum. This generally occurs at a temperature of about $\approx 0.6 T_c$ with a value $\approx 1.8 e_0$. 

We denote the maximum of the latent heat as $L_H$. Its value is provided in the last column of the lower section of Table \ref{table}. The maximum latent heat lies within the narrow range $L_H \approx 29.2-31.2$ MeV. From the previous discussion, one might have expected the associated value to correlate with that of the saturation energy. We find, however, that both UNEDF1 and UNEDF2, which fix the saturation energy, provide an error bar of the same order as UNEDF0, for which $e_0$ was allowed to change in the fitting procedure. 

\begin{figure}
\begin{center}
\includegraphics[width=0.7\linewidth]{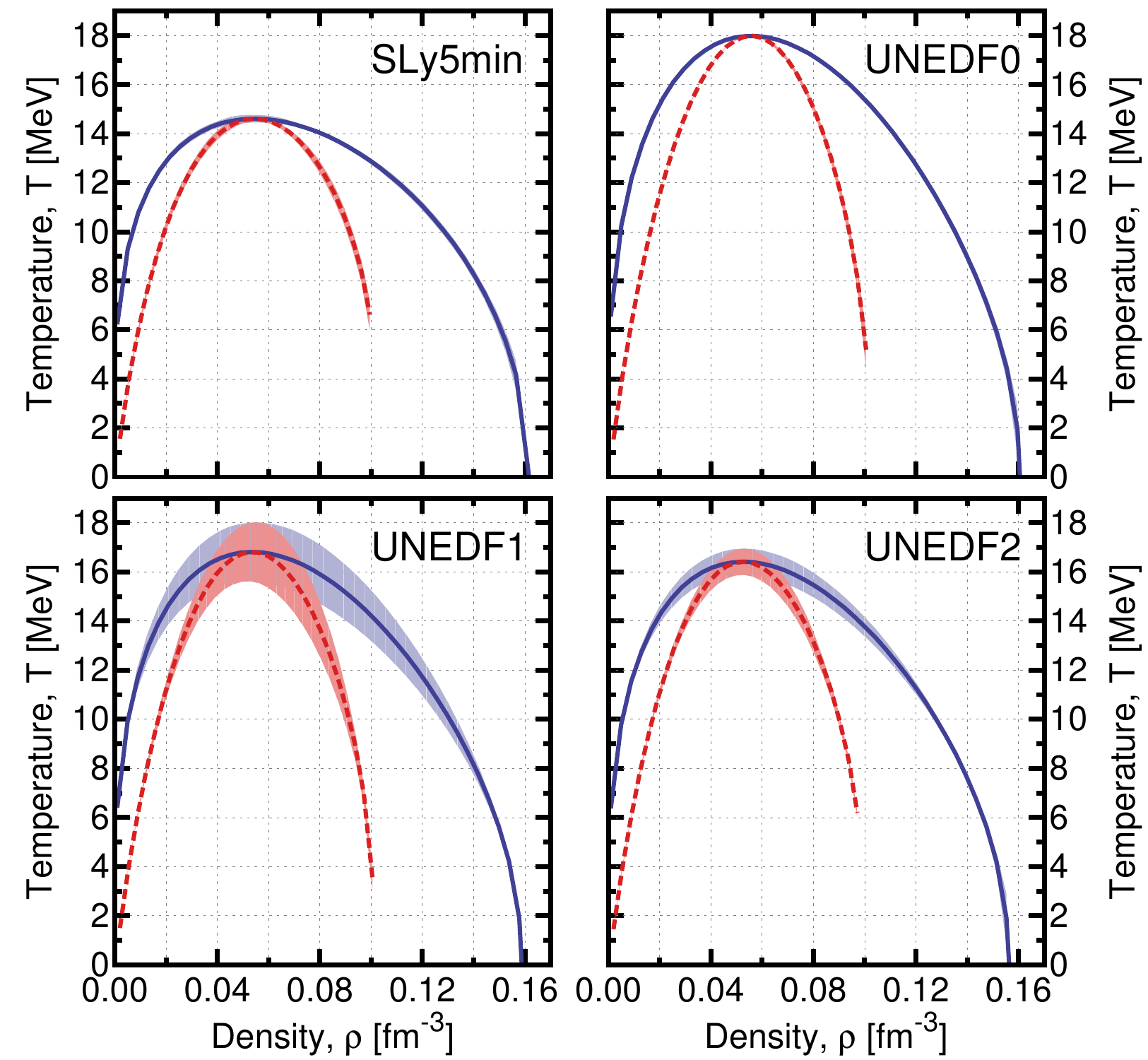}
\caption{(Color online) Coexistence (solid line) and spinodal (dashed line) lines in the temperature-density plane, defining the phase diagram of the liquid-gas phase transition. The bands correspond to $1 \sigma$ errors propagated as described in the text. The meeting point between the coexistence and spinodal line is the critical point. }
\label{fig:coex_spin}
\end{center}
\end{figure}

In the temperature-density plane, two lines determine the phase diagram of the liquid-gas phase transition. On the one hand, the coexistence line is obtained by imposing the Maxwell conditions on the chemical potential to find a set of gas and liquid densities, $\rho_g(T)$ and $\rho_l(T)$. This correspond to an equilibrium configuration,  formed by a certain fraction of both gas and liquid. On the other hand, the spinodal line encompasses the area where the system is mechanically unstable and hence $\partial_\rho p < 0$. The coexistence and the spinodal line coincide at the critical point. In addition to propagating errors for a single quantity, one can also propagate errors for these sets of points. For the coexistence and spinodal lines, for instance, we have propagated the errors in the temperatures for a fixed subset of densities. We present the results of this propagation exercise in \Fref{fig:coex_spin}. 

The critical point and the associated statistical errors can be easily visualised. The coexistence line encloses a wider region of the $\rho-T$ region and has an asymmetric bell shape, as expected from the Guggenheim parametrization of the liquid-gas transition in fluids \cite{Rios10}. The spinodal line encloses a narrow temperature-density region. The mean-field approximation will fail within this region due to the absence of fluctuations \cite{Colonna2002}. For practical reasons, we cut the calculation of the spinodal line at $\rho=0.10$ fm$^{-3}$, but note that the line should extend to larger densities. 

In addition to the central values for the phase diagram, we obtain error bands for the coexistence and spinodal lines. The width of the error bands follows the trends already observed in \Fref{fig:EoS}. SLy5min and UNEDF0 have rather small, almost temperature-independent errors and hence the coexistence and spinodal lines are very well defined - to within the width of the lines presented in the Figure. In contrast, the larger uncertainties in the UNEDF1 and UNEDF2 fit parameters propagate into their respective phase diagrams. We note, in particular, that the error bands increase with temperature, as one moves away from the well-constrained zero-temperature EoS. Again, we associate the statistical errors in the phase diagram to the varying degree of constraint of the isoscalar effective mass.

\begin{figure}
\begin{center}
\includegraphics[width=0.8\linewidth]{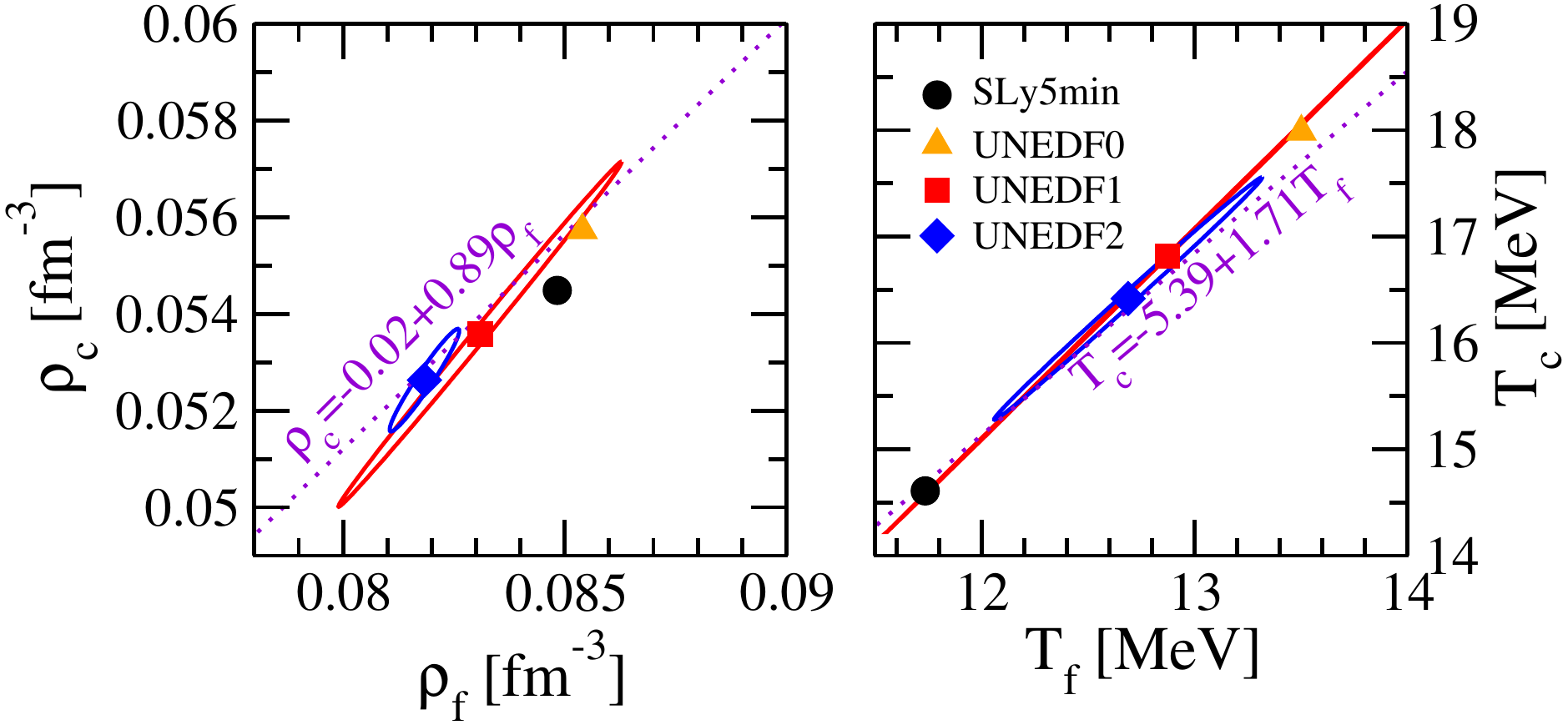}
\caption{(Color online) Left panel: critical density versus flashing density for the 4 EDFs considered here. Right panel: critical temperature versus flashing temperature for the 4 different EDFs considered here.  For UNEDF1 and UNEDF2, we also display $90 \, \%$ confidence ellipses . The dashed lines correspond to the empirical correlation lines identified in Ref.~\cite{Rios10}.}
\label{fig:coex_flash}
\end{center}
\end{figure}

In addition to providing error estimates, Gaussian propagation techniques allow for a clean identification of correlations between parameters. In the context of the liquid-gas transition, one of us identified an empirical correlation between the flashing and the critical properties as provided by a wide set of Skyrme and Gogny forces \cite{Rios10}. More specifically, the critical and flashing densities of all the effective interactions fell very close to the empirically determined line $\rho_c = -0.02 + 0.89 \rho_f$. Similarly, by fitting linearly the results of several mean-field calculations, a tight correlation between the critical and flashing temperatures, $T_c = -5.39 + 1.71T_f$, was identified. 

We show the four coexistence densities (temperatures) as a function of the flashing densities (temperatures) in the left (right) panels of \Fref{fig:coex_flash}. The four EDF results fall close to the empirically determined linear correlations. Physically, these strong correlations indicate that the flashing and critical points are tightly coupled. As a matter of fact, both points are connected by the spinodal line and there seems to be little freedom to change them independently. If this explanation is actually correct, however, one would expect the correlations to appear not only for different Skyrme parametrizations, but also within a single Skyrme parametrization after error propagation. 

We therefore show the $90 \, \%$ confidence ellipsoids of the flashing and critical densities (temperatures) associated to UNEDF1 and UNEDF2 with solid lines in \Fref{fig:coex_flash}. The two densities are extremely well correlated, with off-diagonal normalised Pearson product-moment correlations close to one, $c_{\rho_c,\rho_f} > 0.9998$. Similarly, the temperatures are off-diagonally correlated, with $c_{T_c,T_f} > 0.997$. All in all, this results in extremely thin and alienated confidence ellipses, as can be clearly observed in \Fref{fig:coex_flash}. Perhaps more interestingly, the slopes parallel to the ellipsoids in the $\rho_f-\rho_c$ and the $T_f-T_c$ planes follow closely those derived from the empirical analysis of several EDFs. This demonstrates at once that the connection between flashing and critical points is physically motivated, rather than dependent on the EDF parametrization.

\begin{figure}
\begin{center}
\includegraphics[width=0.44\linewidth]{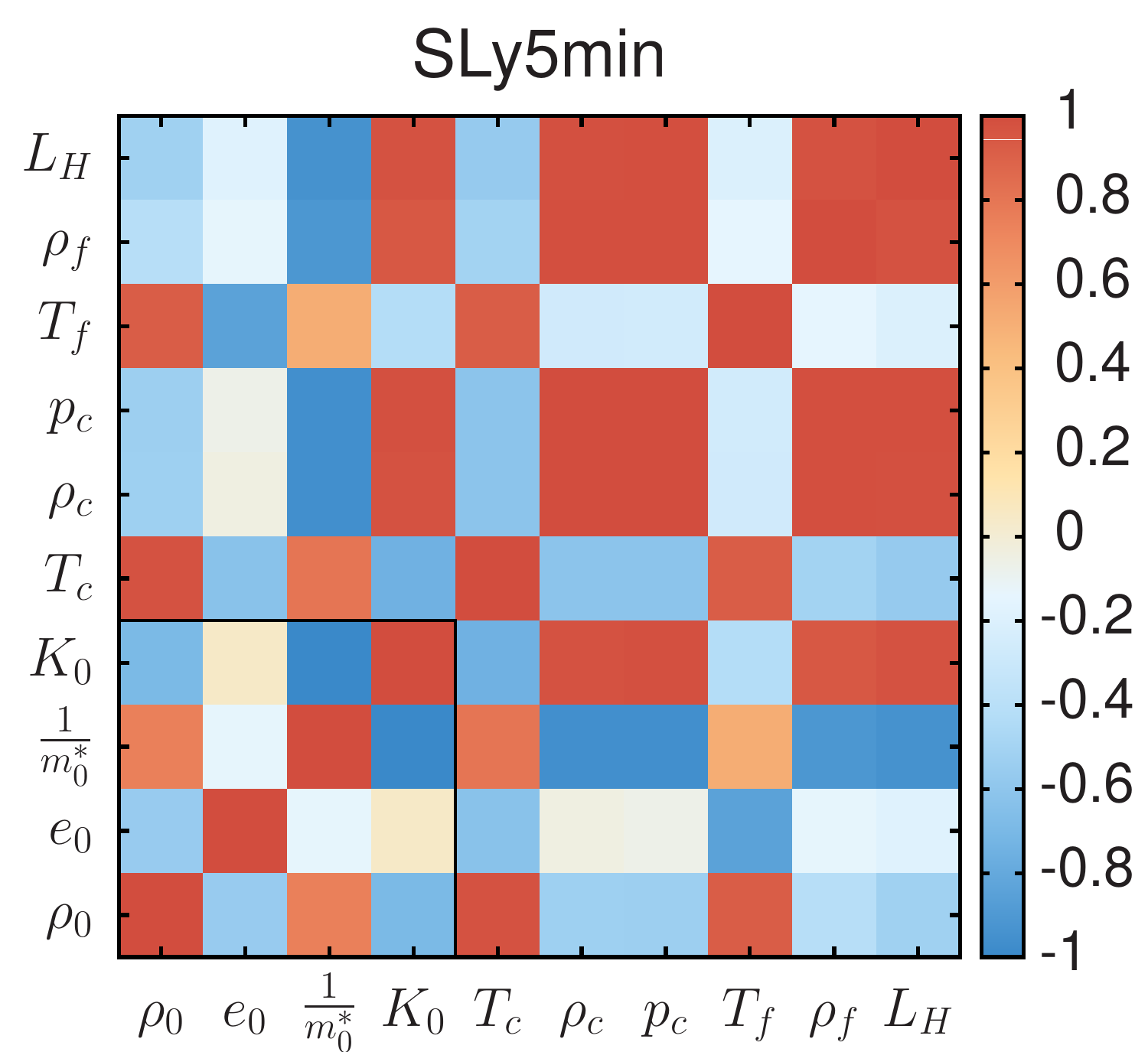} \hskip0.2cm
\includegraphics[width=0.44\linewidth]{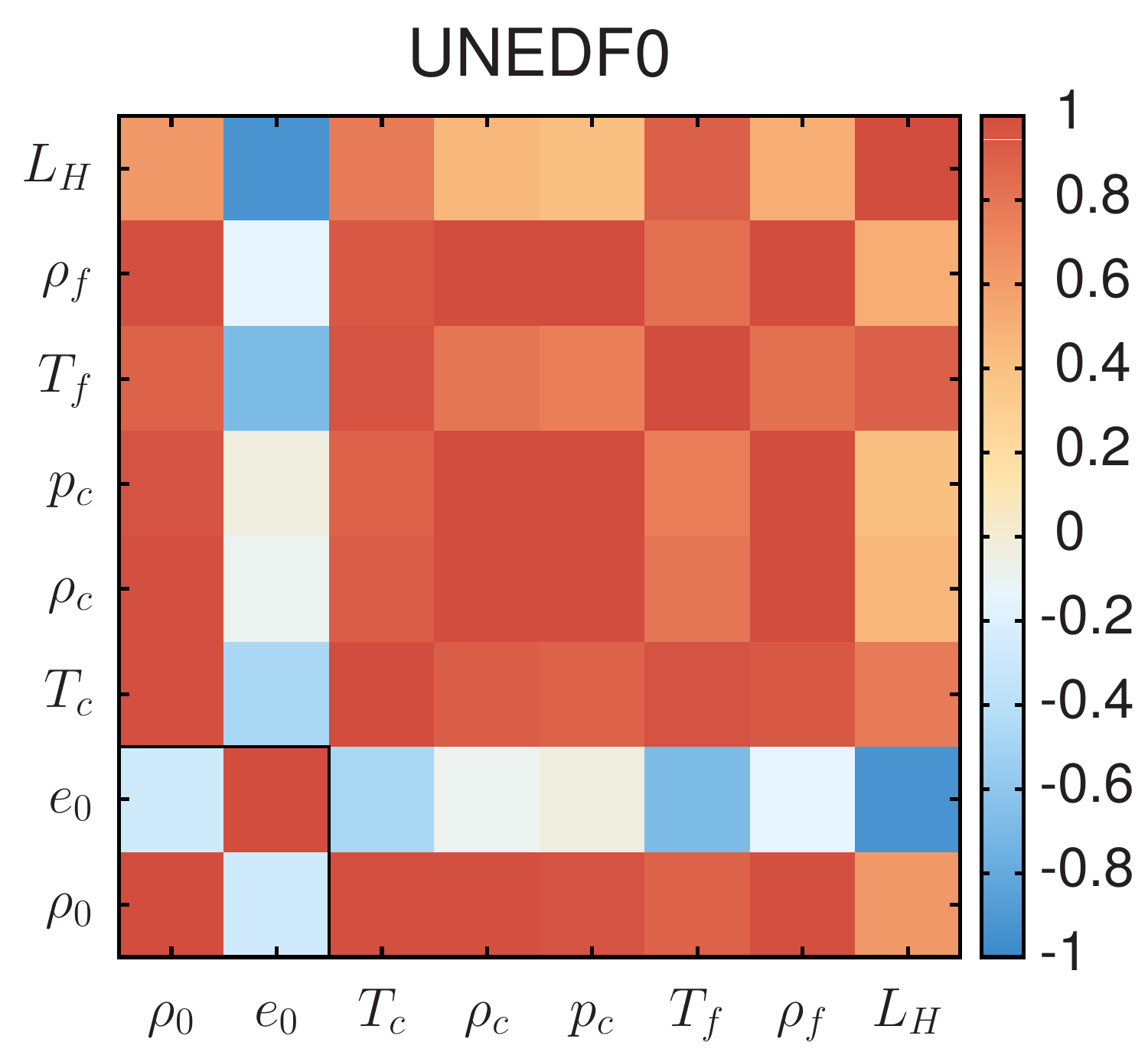}
\includegraphics[width=0.44\linewidth]{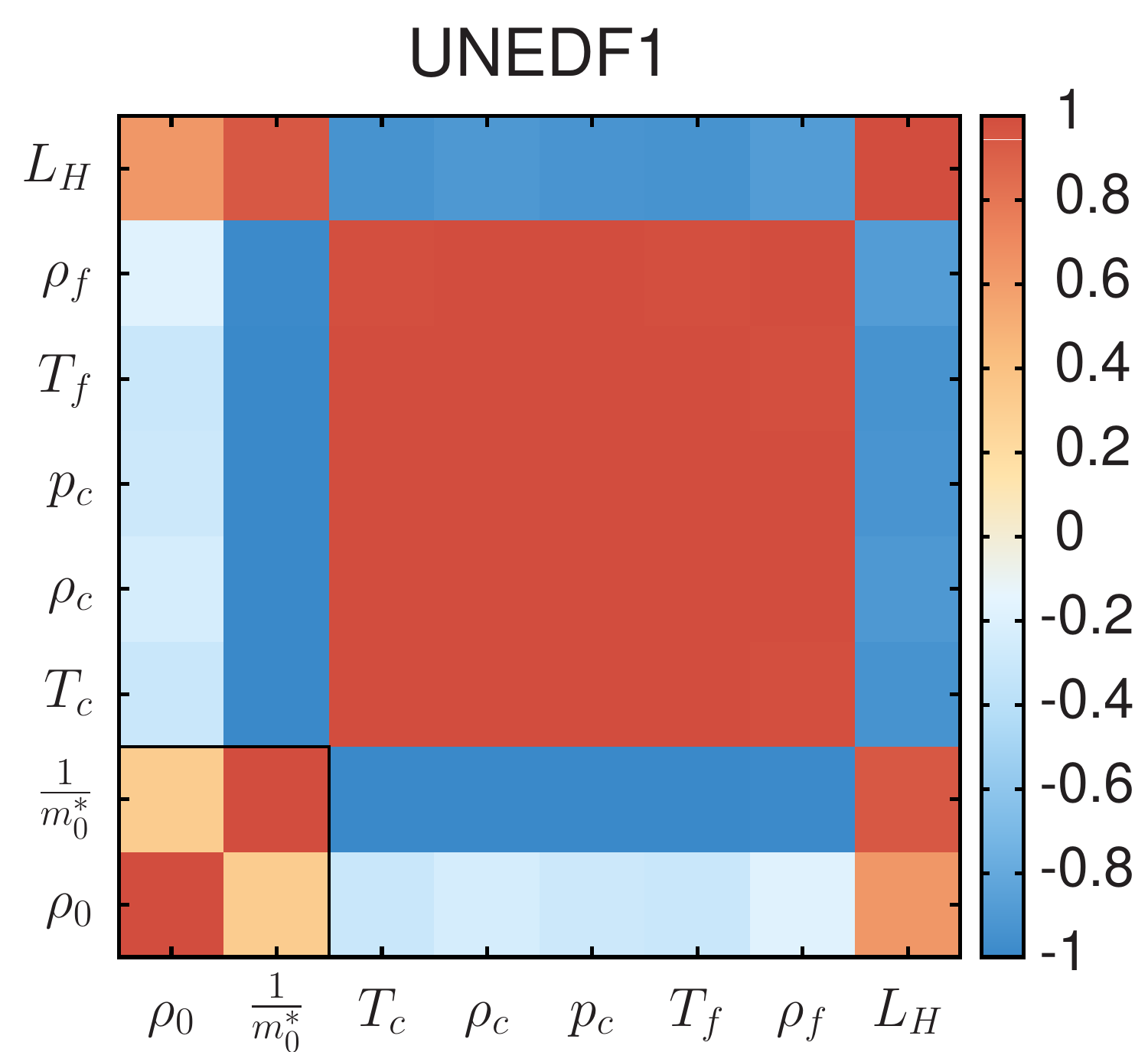} \hskip0.2cm
\includegraphics[width=0.44\linewidth]{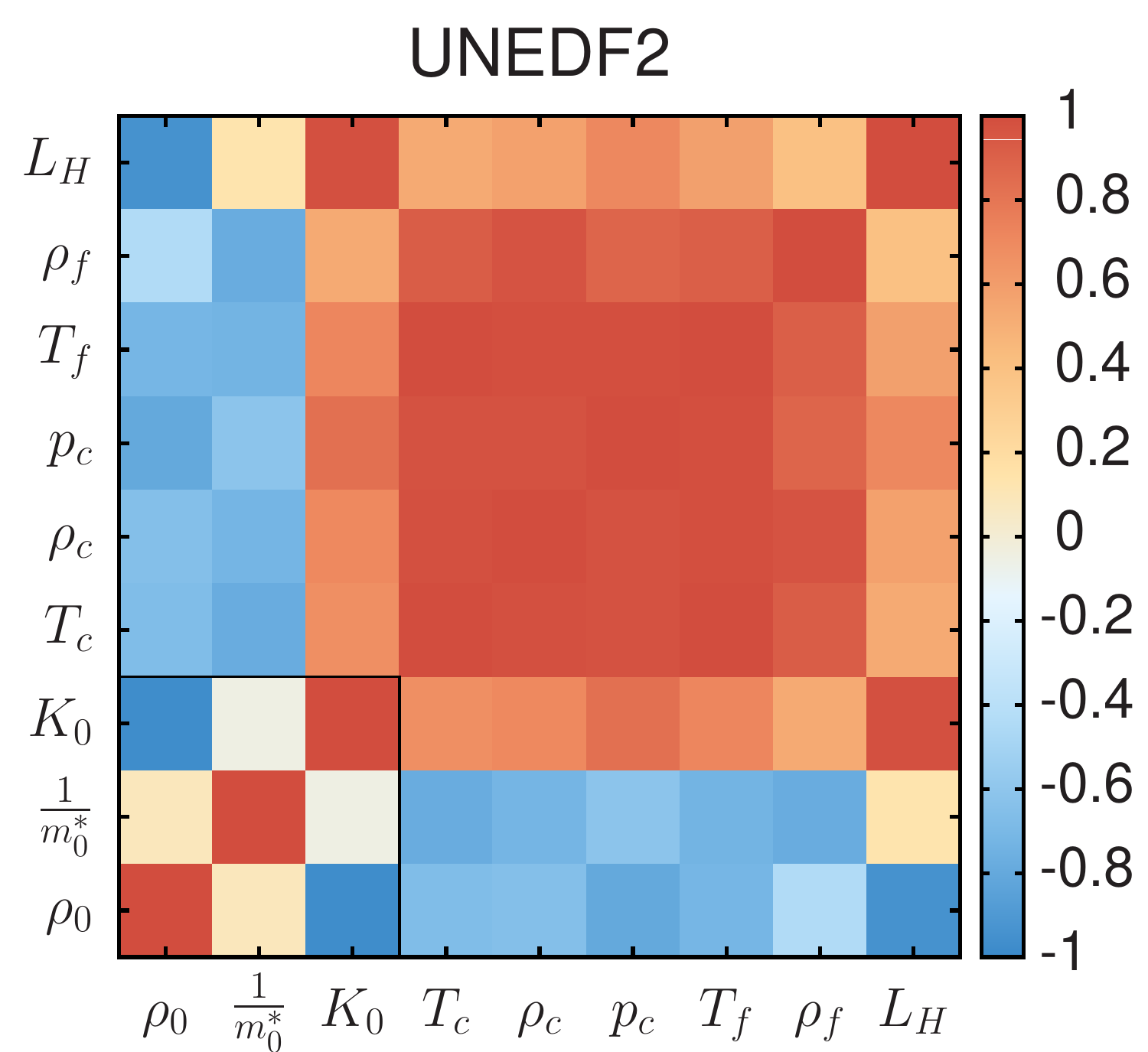}
\caption{(Color online) Pearson product-moment correlation coefficient matrix (colour coded) predicted by the covariance analysis based on SLy5min (top left panel), UNEDF0 (top right panel), UNEDF1 (bottom left panel) and UNEDF2 (bottom right panel). The original correlations between zero-temperature, saturation properties are displayed within a small square in the low, left end of each panel. }
\label{fig:covariances}
\end{center}
\end{figure}

One can extend the correlation analysis to a wider set of observables. We show the Pearson product-moment correlation coefficient matrices for the EDFs under consideration in the panels of \Fref{fig:covariances}. We also provide, for reference, the correlation coefficients of the respective 4 (SLy5min), 2 (UNEDF0-1) and 3 (UNEDF2) free saturation properties in the lower left corner. The features provided by these matrices offer a wealth of interesting results, summarised in the following.

Let us start with SLy5min (upper left panel). Interestingly, the saturation density is tightly correlated to the critical and flashing temperatures which, unsurprisingly, are also tightly correlated. In contrast, the saturation energy hardly correlates with any of the finite temperature properties. The effective mass and the incompressibility show symmetric behaviours when correlating with critical and flashing properties. We note that this is because they are anti-correlated between themselves. In particular, there seems to be a strong anti-correlation between the effective mass and the critical density and pressure. This is a generic behaviour, appearing in all EDFs, suggesting that, once the effective mass has been determined, these critical properties become redundant.

In contrast, the UNEDF0 results show a strong correlation between the saturation density and most finite temperature properties. For this functional, unlike the remaining EDFs, the effective mass is fixed and $\rho_0$ determines to a large extent the finite temperature properties. Because the other functionals have effective masses which are not fixed, it is the latest that correlates more strongly with finite temperature properties. As opposed to SLy5min, we find that most finite temperature properties are strongly correlated among themselves. The exception is the maximum latent heat, which is only strongly linked to the flashing temperature. 

The covariance matrix obtained with UNEDF1 leaves little room for uncorrelated outcomes. As discussed earlier, it appears that the effective mass anti-correlates so strongly with all finite temperature properties that these become immediately entangled with it. All finite temperature quantities are correlated to within $99 \, \%$ or above. These connections are slightly less constrained in UNEDF2 (lower right panel), where one finds that the correlation between saturation properties and finite temperatures is relatively small. Once again, the latent heat is the least correlated finite temperature property. This could in turn be due to its direct relation to the saturation energy. 

All in all, the covariance analysis indicates that the finite temperature critical and flashing properties are strongly correlated among themselves. Since these are not fitted, but rather derived properties, this correlation suggests that they can all be obtained from one another. In other words, there is a certain degree of redundancy in these quantities, even though their relation is not necessarily analytical or easy to find. 

We find a universally (for all functionals) strong correlation between the critical density and pressure. By all measures, moreover, the finite temperature properties are strongly correlated with each other. In other words, the thermodynamics of isospin symmetric systems at different temperatures are to a large extent degenerate. We note that these are not undoubtedly well correlated with a single zero temperature property (except for UNEDF1) and hence the interplay between the uncertainties in finite and zero temperature quantities is not necessarily trivial.

\section{Conclusions and future prospects}

We have presented a preliminary covariance analysis of the thermal properties of nuclear matter. We have concentrated on symmetric nuclear matter and have worked with 4 different functionals (with their respective covariance matrices). We have used an error propagation  analysis based on a $\chi^2$ objective function with a Gaussian likelihood function. A number of relevant results have been obtained. 

First, thanks to covariance matrix techniques, we have been able to quantify the error in the EoS of symmetric matter. With this analysis, we have confirmed the essential role of the isoscalar effective mass in determining the temperature dependence of errors. In particular, for functionals with fixed effective masses, the uncertainties become essentially independent of temperature. In contrast, inaccuracies in the effective masses propagate into the temperature dependence of the bulk properties, so that higher temperature properties become more uncertain. 

Second, we have been able to estimate the statistical errors in the liquid-gas phase transition parameters. We find that all critical properties are very well constrained, to within a few $\%$ even in the most error-prone parametrization, UNEDF1.  The same findings apply to the flashing point and to the maximum of the latent heat in the phase transition, which are constrained to a large extent by the zero-temperature saturation properties. As  a matter of fact, the whole phase diagram for the phase transition is rather well determined. Uncertainties only appear for the largest temperatures in those parametrizations that have non-negligible statistical errors in the effective mass. 

Finally, we have looked at the correlations between different parameters. We have confirmed the already known correlation between flashing and critical points by computing the covariance ellipses for these properties. The results lie very close to the empirical correlations discussed in Ref.~\cite{Rios10}. We have also found that all finite temperature properties show a very strong correlation among themselves. This indicates that they are somewhat redundant, in the sense that they must depend on the same underlying physical properties. 

We foresee a few possibilities for future research based on these results. To begin with, the symmetric matter EDF is so simple that one could investigate error propagation without the Gaussian approximation, using Markov chain Monte Carlo approaches. The calculation of the critical exponents in the self-consistent mean-field theory has already been achieved \cite{Rios10}. A check on the independence of these exponents to the underlying parametrization errors would be of interest. Finally, a natural extension of this approach is the study of the phase transition in the isovector channel. In particular, the important issue of what observables (if any) correlates with the temperature dependence of the symmetry energy and its slope is relevant for the evolution of neutron stars as well as for low energy heavy ion phenomenology  \cite{tsang01,sfienti09}. 

\ack

This work is supported by STFC through Grants ST/I005528/1 and ST/J000051/1. 

\section*{References}

\bibliographystyle{unsrt}
\bibliography{biblio}

\end{document}